# Particle Physics as a way to bring different cultures to work together in Science

G. Mikenberg

Dpt. of Particle Physics and Astrophysics.

The Weizmann Institute of Science

Rehovot, Israel

(e-mail: George.Mikenberg@cern.ch)

Abstract

Science has traditionally played an important role in sharing knowledge among people. Particle Physics, with its large experiments, has shown that one not only can share the knowledge among different cultures, but that one can also work together to achieve this knowledge. The present article gives a few examples where this has been possible among people that are sometimes in conflict situations.

1. Introduction

Science has traditionally played a role of universality, where knowledge and experience where shared through mankind. Particle Physics, through its large and complex experiments, it has provided the possibility for people of different cultures, to not only improve our knowledge, but also to work together, in a common environment, to be able to improve our understanding of matter. This required large infrastructures where collaborative efforts, independent of the people's culture, but based on the common search of knowledge, could be performed. CERN, being the largest of such infrastructure, is the best example.

In the present publication, and based sometimes on partial information, a short report will be given on how CERN allowed for a reintegration of German Physicists into European and World Science, and also served as the starting place to start diplomatic relations between Germany and Israel, after the Holocaust. This will be followed by a set of more personal experiences related to how common work in a series of experiments, and in particular in the ATLAS

MUON Project and its Thin Gap Chamber Detector has allowed to improve contacts and work together for groups from Japan, Palestine, Pakistan and South America. Thus showing that having a well-defined project, at the cutting edge of technology allows people to concentrate on the end product and to forget about possible prejudices.

2. Being German at CERN was not easy in its beginning

Bringing Europe to play a leading role in Science was not an easy task after the end of the Second World War, not only due to the fact that many of the leading scientist had left Europe, but also due to the personal animosity that follows such a conflict, which makes it hard for people to work together. CERN opened the way for German scientists and engineers to become integrated into a large common European scientific project. This was not such an easy task, and from private conversations, it became clear that some of the German colleagues, had a hard time to get accepted even two decades after the end of the war.

One particular German Scientist who played a very important role in this integration was Prof. Wolfgang Gentner [1-2], who through his long time contacts with Joliot-Curie before the war, and through his help in trying to get the Joliot-Curie Cyclotron into operation during the war, as well as his role in freeing him from German prison several times during the war, had an excellent contact to French Physicists. As such, and also due to his experience, he was put in charge of building the first CERN accelerator, the 600 MeV Synchro-Cyclotron, followed by participating in setting the Physics Programme for the 28 GeV CERN PS, and then became CERN Research Director, while being also the Director of the Max Planck Institute (M.P.I.) in Heidelberg.

The strong influence of Prof. Gentner in the early scientific and technical life of CERN, also brought many German Scientists and Engineers to be formed at CERN, which then allowed to create a generation of Accelerator Physicists first at Bonn and then at DESY. Although this early participation at CERN had some integration difficulties, one can conclude that:

> *Having well defined projects at the cutting edge of technology allows people to concentrate on the end project and to forget about their prejudices.*

3. From the Holocaust to start working together for science.

Although the Holocaust was very present in both Israeli and German minds, also political issues had an important role that made rapprochement between the two countries very difficult. In particular, the 1952 Luxemburg Agreement on compensations, was met with strong opposition in both Israeli and German parliaments[3]:

> In Israel, it was called a "pact with the devil", and no cultural relations, except under special approval, was included as part of the legislation.

> In Germany, due to the worry that Arab Countries could recognize East Germany (Hallstein Doctrine), Adenauer had to get votes from the opposition to pass the legislation.

However science, or the search for knowledge are stronger than hate, and although no cultural (or scientific) relations were allowed, the first meeting to collaborate in science took place at the CERN cafeteria.

In 1957 Prof. Gentner (then CERN Research Director) and Prof. de Shalit (then head of the Physics Department at the Weizmann Institute and later its President) met at the CERN cafeteria to discuss possible collaborations between Israeli and German scientists. While the motivation on the German side was to regain the respect of the international scientific community, for the Israeli side it is not so clear, but one can infer that scientist must cooperate, even under the background of national tragedies.

To make it happened, one needs the highest levels: Adenauer and Ben-Gurion:

> Adenauer was contacted by D. Heineman (co-founder of AEG); quoting Adenauer's memories: "As major of Cologne, I had many friends. Heineman and Prof. Kraus (both Jewish), were the only ones who helped me when I was remove from office".

> Ben-Gurion was contacted by de Shalit and the ex-Foreign Minister Aba Eben.

In December 1959 an official delegation from the Max Planck Gesellshaft (MPG), headed by Otto Hahn came for a 10 day visit to the Weizmann Institute and a scientific exchange program started in 1962.

Since 12 May 1965 Germany and Israel have stablished diplomatic relations.

It took a long time for Israeli scientists to start collaborating with German scientists in common projects. I was the first Israeli to come to DESY to work in a common HEP Experiment and I could not understand how we can do science without each other.

4. Japan-Israel Collaboration and how to share responsibilities

Japanese groups under the leadership of Prof. Koshiba have started activities in European Experimental High Energy Physics in the late 60's, first at Novosibirsk and then at DESY, while the Israeli groups started activities at DESY in early 1977. Both groups worked together in the DASP Experiment, but then moved to different Experiments at the PETRA accelerator; Japanese groups in JADE, Weizmann group in TASSO and Tel-Aviv group in PLUTO.

Although being in different experiments, contact was always kept, and in particular, the Japanese colleagues were instrumental, due to their experience in calibration light sources for electromagnetic detectors, in finding the right solution for the calibration of the Hadron Arm Shower Counters of the TASSO Experiment, which was one of the first detectors to use the wavelength shifter technology in scintillation detectors.

The same groups from Japan and Israel joined a common experiment at the LEP collider: OPAL. The Japanese groups developed and build the Barrel Electromagnetic calorimeter, while the Israeli groups developed and build a new type of detector: Thin Gap Chambers [4] (TGC) that were used as the forward hadron calorimeter of the OPAL Experiment (this is described in more detail in the next section). But both groups also played a very important role in keeping the full experiment in running condition and analysing its data. Furthermore, a strong common thrust was achieved from the fact that members of both groups were always present when any problem arised in their respective detector, showing a clear level of responsibility throughout the 11 years that they worked together. It was this mutual appreciation on keeping with their responsibilities that was crucial for the two groups to decide to embark together in a common project for the LHC: the ATLAS End Cap μ trigger[5]. This was not a trivial decision at that point, since most of the Japanese groups were involved in activities in an American project, the SSC. However, the two groups felt that they could thrust each other, to bring such a project to completion, which leads to the following conclusion:

*Mutual appreciation and responsibility are a crucial element for the success of a common scientific enterprise.*

The common project worked well, and both sides took their responsibilities, and the 3,600 TGC detectors with their electronics have been operational at the 99% since 2008.

The Israeli contribution to the OPAL Experiment was an important motivation for the CERN management to suggest that Israel should formalize its relations with CERN, by becoming the first Paying Observer State to the CERN Council and setting the rules for such a procedure. This status was then achieved by Japan a few years later, and the very strong contact between the two groups was crucial in exchanging experiences to make such step a success.

5. Development construction and maintenance of TGC detectors

As mentioned in the introduction, having a well-defined project provides an excellent way for scientists from different cultural backgrounds to work together and forget about any possible prejudices. The construction of various projects based on the TGC Technology has allowed to create bridges between Israeli, Japanese, Chinese, Pakistani and Chilean physicists, engineers and technicians.

The TGC technology was born from the need to produce a thin wire chamber detector to operate as a calorimeter in the pole-tips of the OPAL magnet. Such requirements were due to the need of having a very uniform magnetic field in the tracking detector, combined with a very low stray field in the region of the photomultipliers from the electromagnetic calorimeter, which demanded detectors that could fit into gaps of less than 10mm between the Fe plates of the calorimeter.

In this period, new ideas of developing thin wire chamber detector started to appear in the CERN-Charpak group, mainly pushed by S. Majewski. He was invited to spend a few weeks at the Weizmann Institute, and within two weeks (and using the infrastructure of the laboratory from A. Breskin), a credible detector structure was developed, together with a credible gas mixture, which gave a stable operation at high gain, in a quasi-saturated mode[6].

This was a nice success, however there is an enormous amount of nice ideas in particle detector technologies that were never converted into real devices. The main problem resides on the fact that for a real experiment, one is not necessarily interested in all the properties of a measuring device and its limits, but rather in constructing a device that can be manufactured with available manpower and existing materials; furthermore, it should also respond to the experimental needs and should be affordable. Finally, the need to have a working devise for periods that last for 10's of years, leads to a level of conservatism in operating the devise that does not profit from the best of its properties.

For this reason, it is rare that the same people that develop a new type of measuring devise, are the ones that are involved in its optimization and utilization in a large experiment.

400 large area TGC detectors were constructed and installed in the OPAL Experiment as its Pole-Tip Hadron calorimeter and End-Cap Electromagnetic pre-shower. They operated well during a period of 11 years, however, its operation required a lot of care and dedication. This sense of responsibility for the operation of a detector is something that was common for both the Japanese and Israeli groups, which were always present when any problem with their respective detector (Barrel Electromagnetic Calorimeter in the case of Japan, Pole-Tip Hadron Calorimeter for Israel) component arose (see Fig. 1).

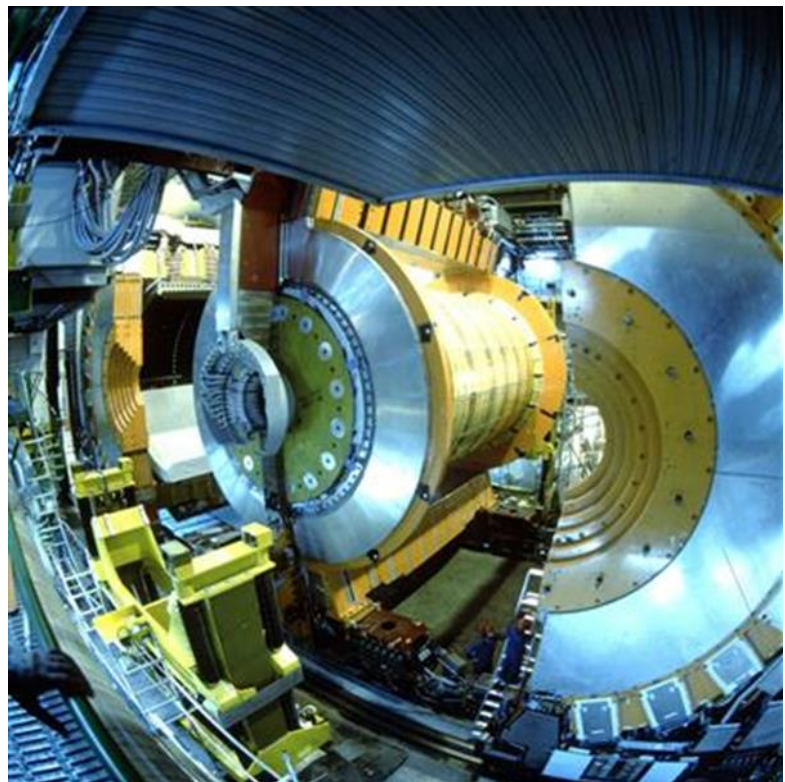

Figure 1: Open view of the OPAL Experiment showing the back of the Pole-Tip Calorimeter (green) and the Barrel Electromagnetic Calorimeter (black and white) in the background.

This sense of mutual responsibility was crucial for the next step of deciding to join forces for the next project, the MUON End-Cap Trigger of the ATLAS Experiment.

The common project worked very well and both sides took their responsibilities (see Fig. 2). The good complementarity between the two communities, with excellent ideas and motivated graduate students in Japan[7–8] and good engineering and technicians in Israel, as well as a lot of common test beams mainly in Japan, allowed to create a real collaboration. Based on the previous experience, the TGC detectors are operated under more conservative conditions as before and both groups make sure that someone knowledgeable is available when any problem arises during the operation.

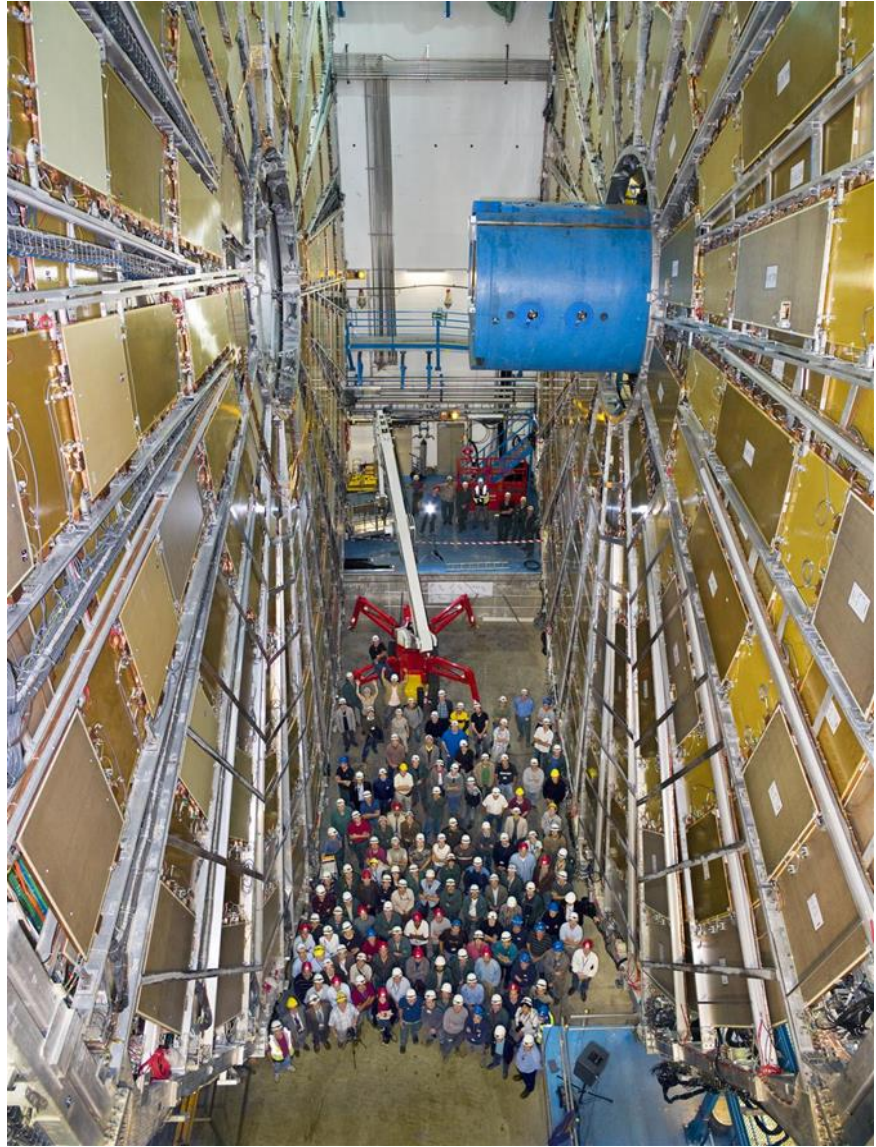

Figure 2: completion of the Big-Wheels project showing the TGC detectors constructed in a collaborative effort between Israel-Japan and China.

6. How Israelis and Palestinians can work together in Science

Although Israel became an Observer State to the CERN Council in 1991, Israeli participation in the CERN Summer Student Program, via its CERN contribution started only in 2002.

In 2005 it was decided that this contribution could also be used to support Palestinian Summer Students.

CERN, being a neutral ground, provides both sides an excellent opportunity to work and celebrate together without the feeling of motherhood.

*To ensure a fruitful Scientific and Cultural collaboration, one should avoid the feeling that one side is being patronized. CERN being a neutral ground, it provides an excellent example to help achieve this.*

And not only could Palestinian and Israeli students could work on common projects, but since they have similar cultures, at least food-wise, a common party for all summer students took place at CERN. As described by the Summer Students themselves, in the CERN Bulletin: "the decision to organize our own party was taken during the Italian party. Besides showing that the reality is not what you see in the news, we wanted people in Europe to experience a different kind of party. With local music and food such as hummus, labane, pita bread and mahalabie for dessert that we made ourselves, the party was indeed different from all others. The party had more gimmicks such as writing all the signs in English from right to left, or a place where people could practice writing in Arabic and Hebrew, and a screen where we projected animation from Israel and Arab belly dancing". Some of the scenes of the party can be seen in Fig. 3.

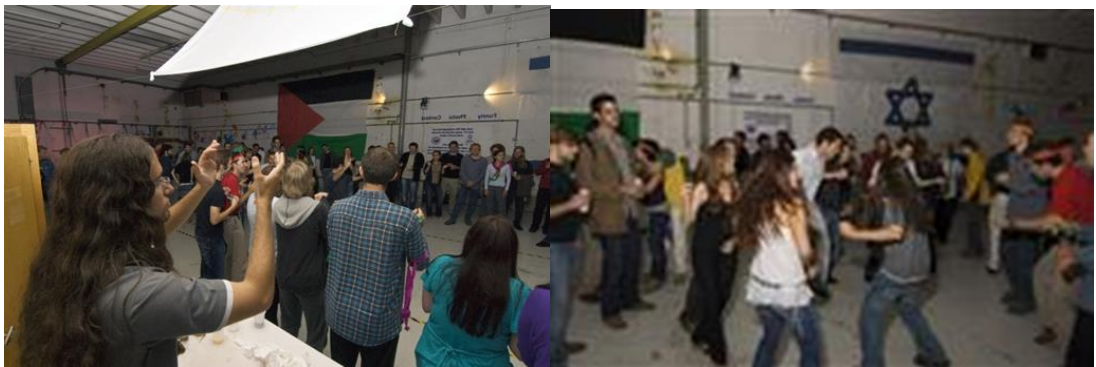

Figure 3: some scenes of the common Israeli Palestinian Summer Student Party, organized at CERN.

7. How Israelis and Pakistanis can work together for Science

Scientists are used to work together independent of their cultural and religious backgrounds due to their motivation to improve our knowledge.

It was not at all clear that this could be possible for groups of engineers and technicians, where this motivation, except for the economic side, is not necessarily present.

With Professor H. Hoorani from Pakistan it was decided that it was worth a trial for one of the big projects of the ATLAS Experiment "the ATLAS Big Wheels":

- The chambers were constructed in Israel, Japan and China
- The electronics were developed and build in Japan
- The support structures (precise large Al structures) were designed in Russia and CERN and made in Israel.
- The jigs and tooling were designed and made in Pakistan

During 3 years 20 engineers and technicians from Pakistan worked together with 20 Engineers and technicians from Israel to put together the project (see Fig. 4).

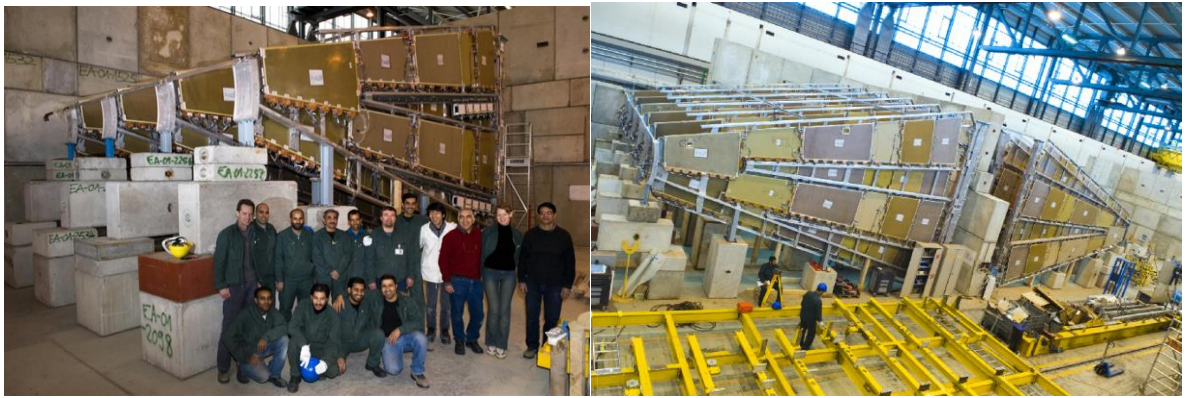

Figure 4: part of the Israel-Japan-China-Pakistan groups putting together the Modules of the ATLAS Big-Wheels and overview of the assembled modules and jigs

They did not only work together, but they also had fun together (see Fig. 5).

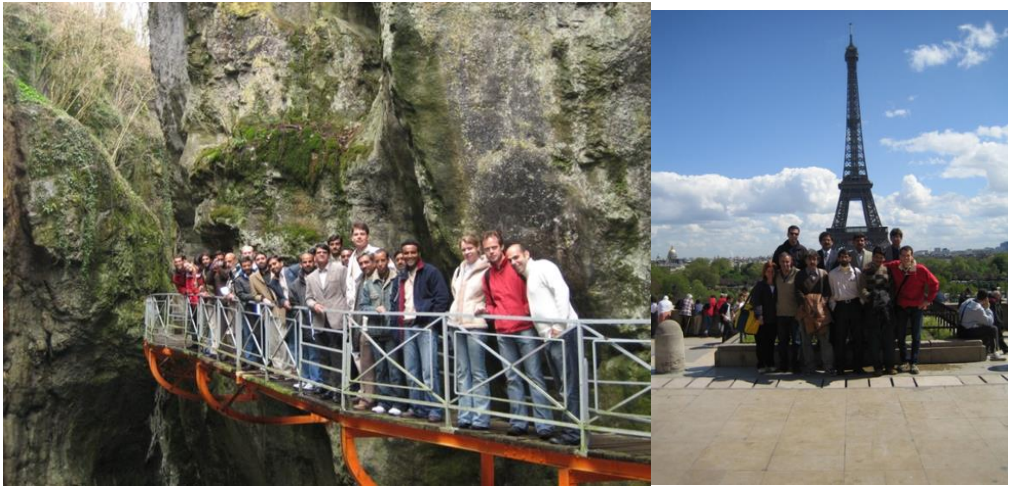

Figure 5: Pakistani-Israeli technical groups during outs to Annecy and Paris.

Most of these people were not particularly interested in the excitement of trying to find the Higgs Boson, but rather very proud of working at the forefront of technology, which leads to:

*Working on projects at the cutting edge of technology allows people to forget about their mutual prejudices, learn to respect each other and feel proud of their common achievement*

It is very hard to make new developments without a clear applicable goal.

CERN, by providing projects for basic research, that are at the edge of the technological possible, allows one to improve/develop technologies and test them in a real environment.

With experiments that include 100M single detector elements that operate with~1% failure, this is tremendous test for developments.

8. Involving South American groups to play a critical role in the large CERN Experiments

Experimental High Energy Physics groups from Latin America have been involved in many experiments for more than 40 years. They have been working at FNAL, BNL and CERN with good success, but mainly based on individual contacts, and not in a coherent form. Most of the hardware contributions being mainly on electronics and scintillator detectors, but not on producing central components of experiments, with its related responsibility.

In the case of Chile, the two main experimental groups (Pontificia Universidad Catolica de Chile, PUC and Universidad Tecnica Federico Santa Maria, UTFSM) joined forces to construct a large portion of the future ATLAS upgrade, called New Small Wheels, to improve the trigger of MUONs in the forward direction. This will also be based on the Thin Gap Chambers technology, and the two groups have been the first ones, within a large international collaboration, to produce the first full size final module of the detector (see Fig. 6). Having such a responsible role, have led the groups to a very high motivation, which was the main reason for this success.

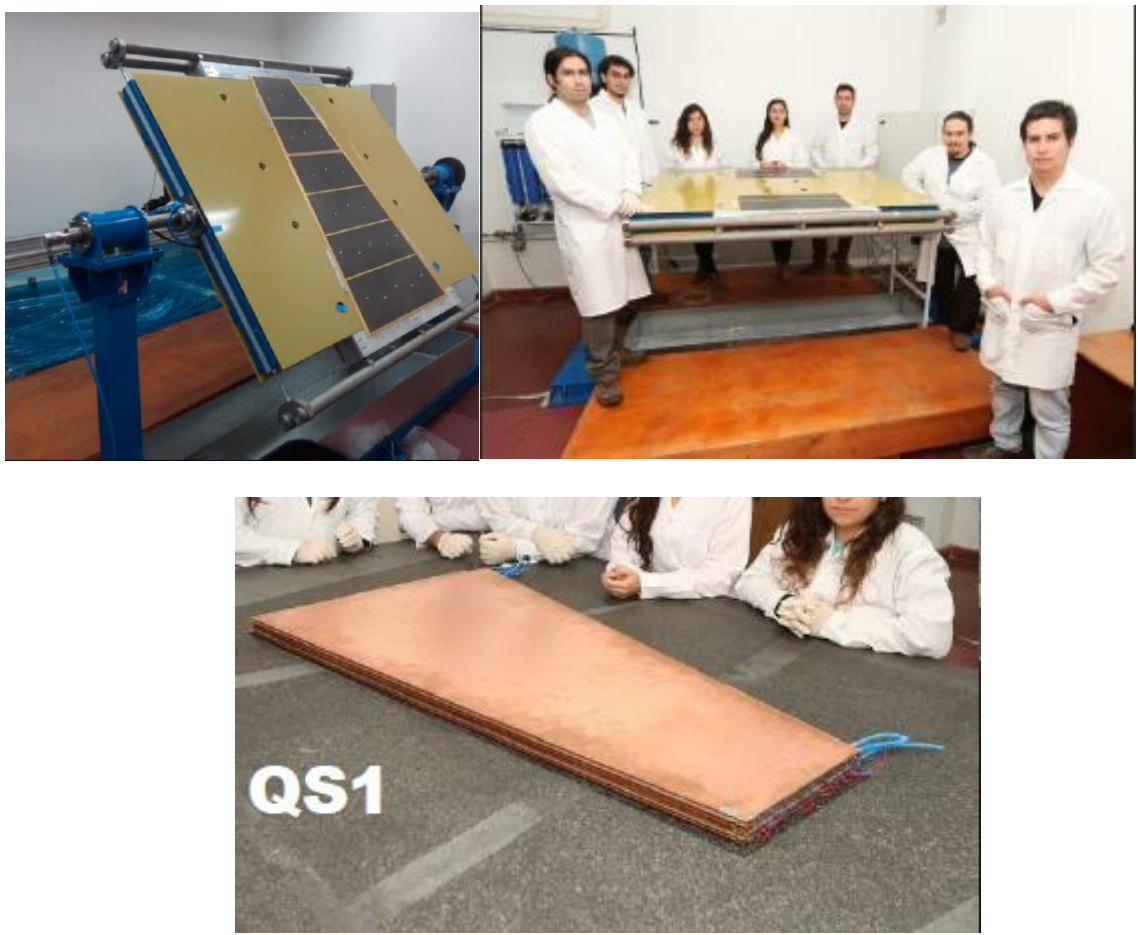

Figure 6: parts of the infrastructure of UTFSM during the winding and preparing the assembly of a quadruplet of the final first production module of the future ATLAS upgrade, shown in the bottom picture.

9. Conclusions

Having well defined projects at the cutting edge of technology allows people to concentrate on the end product and to forget about their prejudices.

CERN provides the possibility to materialize such projects, at the forefront of technology. This allows people to learn to respect each other and feel proud of their common achievement.

CERN, by providing projects for basic research, that are at the edge of the technological possible, allows to improve/develop technologies and test them in a real environment

Mutual appreciation and responsibility is a crucial element for the success of a common scientific enterprise.

To ensure a fruitful scientific and cultural collaboration, one should avoid the feeling that one side is being patronized. CERN is an excellent example on how to achieve this by being in neutral grounds.

To make an impact in large HEP Experiments, it is crucial for local groups to collaborate among themselves in common projects.

References

1) Charles Wainer, Oral stories, American Institute of Physics (https://www.aip.org/history-programs/niels-bohr-library/oral-histories/5080), November 15th, 1971.
2) Cathryn Carson, Heisenberg in the Atomic Age: Science and the Public Sphere, page 212; Cambridge University Press, 2010.
3) Hanan Bar-On, The Role of Scientists in mitigating international discord; Annals of the New York Academy of Sciences, Volume 866, July 2004.
4) G. Mikenberg, Thin Gap Gas Chambers For Hadronic Calorimetry; . Nucl.Instrum.Meth.A265:223-227,1988.

5) K. Nagai et al; Thin Gap Chambers in ATLAS. Nucl.Instrum.Methods A 384;219-221, 1996.
6) S. Majewski, G. Charpak, A. Breskin, G. Mikenberg; A Thin Multiwire Chamber Operating In The High Multiplication Mode. Nucl.Instrum.Meth.217:265-271,1983.
7) S. Tsuno, T. Kobayashi and B. Ye; Gamma-ray sensitivity of a thin gap chamber. Nucl.Instrum.Methods A 482;667-673, 2002.
8) H. Nanjo et al; Neutron sensitivity of thin gap chambers. Nucl.Instrum.Methods A 543; 441-453, 2005.